\documentclass[twocolumn,showpacs,preprintnumbers,amsmath,amssymb]{revtex4}
\usepackage{graphicx}
\usepackage{dcolumn}
\usepackage{bm}
\usepackage{wasysym}
\usepackage{amsmath}
\usepackage{amssymb}

\newcommand{\aj}{AJ}
\newcommand{\apjs}{ApJS}
\newcommand{\apjl}{ApJL}
\newcommand{\mnras}{MNRAS}

\newcommand{\aap}{A\&A}

\newcommand{\cvir}{c_{vir}}
\newcommand{\rvir}{R_{vir}}
\newcommand{\mvir}{M_{vir}}

\newcommand{\bv}{\bar v_{col}}
\newcommand{\vcol}{v_{col}}

\begin{document}

\title{Galaxy collisions as a mechanism of ultra diffuse galaxy (UDG) formation}
\author{Anton N. Baushev}
\affiliation{Bogoliubov Laboratory of Theoretical Physics, Joint Institute for Nuclear Research,
141980 Dubna,
Moscow Region, Russia\\
Departamento de Astronom\'ia, Universidad de Chile, Casilla 36-D, Correo Central, Santiago,Chile}

\email{baushev@gmail.com}
\date{\today}

\begin{abstract}
We suggest a possible mechanism of ultra diffuse galaxy formation: the UDGs may occur as a result
of a central collision of galaxies. If the galaxies are young and contain a lot of gas, the
collision may kick all the gas off the systems and thus strongly suppress any further star
formation. As a result, the galaxies now have a very low surface brightness and other properties
typical of the ultra diffuse galaxies. We use the Coma cluster (where numerous UDGs were recently
discovered) to test the efficiency of the process. The mechanism works very well and can transform
a significant fraction of the cluster population into ultra diffuse galaxies. The UDGs formed by
the process concentrate towards the center of the cluster, and their globular cluster systems
remain undamaged, in accordance with observational results. The projected surface density of UDGs
in the cluster may help us to recognize the mechanism of UDG formation, or clarify relative
contributions of several possible competitive mechanisms at work.
\end{abstract}


\maketitle

\section{Introduction}

Unusual properties of the ultra diffuse galaxies (hereafter UDGs) has drawn a lot of attention to
these objects. While some properties of them are similar to that of long-known low surface
brightness galaxies (hereafter LSBs), recent observations allow us to separate UDGs among LSBs and
even may give reason to suppose that UDGs form a separate class of objects.

As with LSBs, the UDGs have unusually low central surface brightness $\mu_{g,0}\sim 24-26$~
mag/arcsec$^2$, while their effective radii are $r_e=1.5-4.6$~{kpc}, which is comparable with that
of the Milky Way ($r_e\simeq 3.6$~{kpc}) \cite{koda2015}. However, recent observations
\cite{dokkum2015a} have surprisingly found 47 UDGs in the Coma cluster, while the standard theory
of the LSBs suggests that they hardly can be formed in high-density environments (see, for
instance, \cite{dekelsilk1986}). Meanwhile, \cite{koda2015} found 854 UDGs in the Coma cluster, and
their density grows towards the center. It means that UDGs can occur in dense clusters and even be
numerous there. Their survival in the environment with strong tidal perturbations suggests that
UDGs are highly dark matter-dominated systems.

UDGs are quite red and show no feature. The featurelessness and low brightness makes the mass
measurements very challenging. However, the object VCC 1287 in the Virgo cluster has a rich system
of globular clusters, and the mass of the galaxy was recently estimated as $\sim 8\times 10^{10}
M_\odot$ \cite{beasley2016}, i.e., this is a dwarf. On the contrary, \cite{dragonfly44} estimated
the mass of the Dragonfly 44 galaxy in the Coma cluster as $\sim 10^{12} M_\odot$ and reported
about $\sim 100$ globular clusters surrounding this object. If this estimate is correct, Dragonfly
44 has a giant dark matter component, comparable with that of the Milky Way. From globular cluster
counts, the median UDG halo mass $\sim 1.5\cdot 10^{11} M_\odot$ \cite{vandokkum2017}.

The fact of the presence of many globular clusters around some UDGs seems very noteworthy: whatever
the UDG formation mechanism is, it does not destroy the globular system around the objects.
Originally UDGs were believed to be round objects \cite{loeb2016}, but resent observations suggest
that UDGs are prolate rather than oblate spheroids \cite{burkert2017}.

An extensive literature discussing the origin of the UDGs has evolved. A very interesting
explanation was offered by \cite{loeb2016}: as the surface brightness is believed to depend on the
galaxy spin, the UDGs can be just the most rapidly rotating tail of the dwarf galaxy distribution.
An UDG may occur if the star formation in the young galaxy was interrupted by AGN feedback
\cite{reines2013}, gas stripping \cite{yozin2015}, or too strong feedback from massive star winds
and supernovae \cite{calura2015}.

We will not discuss the applicability the above-listed models. The aim of this short paper is to
suggest another possible mechanism of UDG formation.

\section{The mechanism}

The idea of the mechanism is that UDGs may occur as a result of central collisions of galaxies. The
collision-less components of the galaxies (dark matter and the stars) should penetrate through each
other freely in such a collision, while the gaseous components collide. The collision heats the gas
and kicks it off the galaxies (see the proof in the next section). As a result, we have two
galaxies with relatively unaffected dark matter and stellar components, but with little gas, and a
separated cloud of hot gas between them. Of course, further star formation is strongly suppressed
in the galaxies. The well-known Bullet cluster gives us a striking illustration of the process
\cite{bullet2}, though on much larger scales. Apparently, the parallels between the Bullet cluster
and the cluster formation are imperfect, since the picture of cluster formation is quite
complicated. However, Bullet cluster illustrates how almost all the gas can be removed from the
system.

We can make a simple estimation of the number of UDGs that could be generated in the Coma cluster
by this mechanism. We suppose that the cluster contains $N_1 \sim 700$ giant galaxies of mass
$\gtrsim 10^{11} M_\odot$ and $N_2 \sim 3\cdot 10^3$ galaxies\footnote{This value may seem
overestimated. However, considering that 854 UDGs have just been found in the cluster, $N_2$ hardly
can be much less than  $2\cdot 10^3$, while the results of our estimate are not that sensitive to
$N_2$.} of mass $\gtrsim 10^{10} M_\odot$ \cite{colless1996}. The galaxies have the King's
distribution in space $n(r)\propto \left(1+(r/r_k)^2\right)^{-3/2}$, where $n$ is the space density
of the galaxies, $r_k\simeq 170$~{kpc} (which corresponds to the angular distance $\theta_k=6'.4$)
\cite{king1972}. We assume that the characteristic radius of the galactic dense gas zone is
$r_1=8$~{kpc} for the first above-mentioned group and $r_2=3$~{kpc} for the second one,
respectively, in accordance with the SDSS galaxy mass-size relation \cite{masssize}, and the
characteristic "cross-sections" are $\sigma_1=\pi r^2_1$ and $\sigma_2=\pi r^2_2$. We assume that
the averaged relative speed of the galaxies in the Coma cluster is approximately equal
to\footnote{See the next section for details.} the doubled velocity dispersion $D[v]$ in the
cluster $\bv\simeq 2 D[v]\simeq 2\cdot 1000$~{km/s}. Then the number of the central collisions $k$
in a time interval $t$ is
\begin{equation}
\label{20a1}
 k=\int\frac{1}{2} t \sigma \bv n^2(r)\cdot 4 \pi r^2 dr\simeq 9.2\cdot 10^{-4} \dfrac{t\sigma \bv N^2}{r^3_k}
\end{equation}
Substituting the above-listed parameters and $t=10^9$~years into equation~\ref{20a1}, we obtain
approximately $40$ central collisions between massive galaxies ($M\gtrsim 10^{11} M_\odot$), $\sim
150$ collisions between a massive and a middle-mass galaxy ($M\gtrsim 10^{10} M_\odot$), and $\sim
100$ collisions between middle-mass galaxies each $10^9$~years.

Before proceeding further, we need to clarify several important issues. First of all, we have
neglected several effects in our consideration. We assume that the averaged collision speed $\bv$
of the galaxies is approximately equal to the doubled velocity dispersion $D[v]$ in the Coma
cluster. We may neglect the dispersion variations with radius in our estimate: the dispersion at
$10 r_k$ is only $25\%$ less than in the center, while the galaxy density already decreases $\sim
1000$ times \cite{king1972}. The interrelation between $\bv$ and $D[v]$ depends on the velocity
distribution of galaxies, specifically on its anisotropy (on the other hand, the distribution
affects the mechanism efficiency only through $\vcol$). For the Maxwell one $\bv\simeq 2.26 D[v]$,
but the radial motion typically dominates in clusters, and we may expect a highly anisotropic
velocity distribution. The extent to which $\bv$ is affected by the anisotropy depends on the space
distribution of galaxies in the Coma cluster, as well as on its gravitational field profile. Both
these quantities lack precision now. However, the solution may be found analytically for a much
simpler, but rather similar case: \cite{11} compared the velocity distribution of the dark matter
particles on the Solar System orbit for the isotropic and for the limiting anisotropic cases. It
turned out that $\bv$ is only $1.27$ times higher for the radial case than for the isotropic one.
The factor $1.27$ is within the accuracy of our analysis. On the other hand, $\bv\simeq 2.26 D[v]$
and even higher for the anisotropic case. It means that by the supposition $\bv\simeq 2 D[v]$ we
underestimate the number of galaxy collisions.

When we introduce the characteristic "cross-sections" of the gaseous components $\sigma_1$ and
$\sigma_2$, we imply that the gas in the galaxies forms spherical clouds of radii $r_1$ and $r_2$.
The question of real gas distribution in young galaxies is not quite clear, but it was hardly
spherically symmetric, and some form-factors should be introduced to take into account real galaxy
shapes and their mutual arrangements. A precise calculation of the form-factor requires detailed
information about the gaseous component of the young galaxies, as well as a reliable model of the
cluster formation. For simplicity, we consider $\sigma_1$ and $\sigma_2$ as the quantities already
averaged over all possible geometries.

In deriving equation (\ref{20a1}) we assume that the galaxies move along straight lines. However,
their gravitational attraction pulls the trajectories together, and their pericenter distance is
always less than the impact parameter. Thus the gravitational interaction increases the collision
rate, and the magnitude of this effect is defined by the ratio of the galaxy virial speed
$v_{vir}=\sqrt{G\mvir/\rvir}$ to the collision speed $\vcol$. The ratio is small in our case, and
so is the effect of attraction.

\section{Can the galaxy collision remove the gas from the systems?}

Let us perform a tentative analytical calculation suggesting that a head-on collision of even a
middle-mass galaxy with a massive one very likely removes most of the gas from both the systems,
either by direct kicking it off, or by strong shock heating that just evaporates the gas from the
system. The smaller galaxy necessarily loses a very significant part of the gas component in such a
collision.

Indeed, the typical speed of the collisions ($\bv\sim 2000$~{km/s}) significantly exceeds the
escape speed from the centra of even massive galaxies. This relation appears to be valid for any
galaxy cluster, not only for Coma. If we suppose that the galaxies, as well as the cluster, have
the NFW density profile, their central potential is $\phi_c=-\dfrac{G \mvir\cvir}{\rvir A(\cvir)}$,
where $\mvir$, $\rvir$, $\cvir$ are halo virial mass, radius, and concentration; $A(\cvir) \equiv
\ln(\cvir+1)-\dfrac{\cvir}{\cvir+1}$ \cite{9}. If we introduce the average halo density
$\langle\rho\rangle=\mvir/\frac43 \pi \rvir^3$, we obtain $\mvir/\rvir=(\frac43 \pi
\langle\rho\rangle \mvir^2)^{1/3}$ and
$$
\phi_c=-\frac{G \cvir}{A(\cvir)}\left(\frac43 \pi\right)^\frac13 \langle\rho\rangle^{1/3}
\mvir^{2/3}
$$
Since $\langle\rho\rangle$ and $A(\cvir)$ have only a weak dependence on $\mvir$ \cite{gorbrub2},
$\phi_c\propto\cvir \mvir^{2/3}$ to sufficient accuracy. The galaxy velocity dispersion is mainly
defined by $\phi_c$ of the cluster, concentrations $\cvir$ for galaxies unlikely can exceed more
then ten times the one for the cluster, while the galaxy masses (excepting the central galaxy) are
$2-3$ orders of magnitude less than the cluster mass. Therefore, the typical galaxy collision speed
in the central area of a cluster (where the mechanism under consideration is the most effective)
should significantly exceed the escape speed from the centra of galaxies.

A precise calculation of the gas fraction a galaxy can lose as a result of collision is very
challenging and requires a reliable model of the galactic gaseous component (especially, for high
redshifts), as well as complex magnetohydrodynamic simulations. However, we may obtain quite
reasonable estimations by mere usage of a toy model and the momentum conservation law. Moreover, we
need to clarify what we mean under the 'central galaxy collision'.

The gaseous interstellar medium in galaxies is a very complex object; however, it can be roughly
subdivided into three components \cite{ism}. Let us use the Milky Way galaxy to illustrate. Cold
medium (CM) is composed of separate clouds with the temperature $T\sim 100$~{K}, particle density
$\varrho\sim (20-50)$~{cm$^{-3}$}, and scale height $h\sim 200$~{pc}; it is the only component
where the star formation may occur. Warm medium (WM) has the temperature $T\sim 10^4$~{K}, particle
density $\varrho\sim (0.2-0.5)$~{cm$^{-3}$}, and scale height $h\sim 300$~{pc}. The coronal gas has
the temperature $T\sim 10^6-10^7$~{K}, scale height $h\sim 3000$~{pc}, and the particle density
that rapidly drops with radius from $10^{-2}$~{cm$^{-3}$} to $10^{-4}$~{cm$^{-3}$}. It is
significant that, despite its low density, the interstellar medium (ISM) is not in vacuum state:
the free path of the particles is typically much shorter than the area size.

Let us take up a collision of two galaxies with velocities $v_1$ and $v_2$ ($v_1+v_2=\vcol$).
Consider a narrow cylinder of cross section $S$ and with the axis parallel to $\vcol$. The momenta
of the ISM inside the cylinder are $\nu_1 S v_1$ and $-\nu_2 S v_2$. Here $\nu$ is the projecting
density $\nu= m_a \int\varrho dl$; the integral is calculated along the line parallel to $\vcol$
across each galaxy, $m_a$ is the average atom mass. The overlapping parts of the ISM undergo a
non-elastic collision, and a shock wave occurs between them, but the net momentum conserves.
Therefore, the direction and speed of the substance after the collision is determined by the sum of
the initial momenta, which mainly depends on the relationship among the projecting densities $\nu$
of the galaxies at the given point.

The foregoing parameters of the ISM show that if the axis of the cylinder under consideration
passes through the CM, it makes the main contribution into $\nu$: though the cold clouds are
smaller, their density is much higher. It is not surprising: the ISM is (as a very rough
approximation) in hydrodynamical equilibrium, i.e., the pressure in it is equal. Therefore, the two
orders of magnitude hotter WM is two orders of magnitude less dense. The gaseous corona is of large
extent, but its density is even lower and rapidly falls with distance. Therefore, the projected
density $\nu$ is large if the cylinder crosses CM regions of the galaxy, significantly lower if it
does not, and rapidly drops as the cylinder axis recedes from the galaxy center.

There are two possible geometries of galaxy collision. First, when the central dense areas of the
galaxies (containing CM) pass only through the hot coronas of the other galaxy. Since the
projecting density $\nu$ and the momentum flux density ($\sim m_a \varrho \vcol^2$) in the corona
are incomparably lower, the central galaxy areas pass through them almost without any resistance.
Obviously, such a collision cannot lead to an UDG formation, being therefore of no interest for us.

The second possibility is a collision when CM components of the galaxies significantly overlap
(this is exactly the case that we name 'a central galaxy collision' in this paper). Two regions
with comparable density and $\nu$ collide in this case. We want to demonstrate that in principle
all the gaseous component can be removed from both the systems in this instance.

Let us take up a collision of two galaxies of masses $M_1$ and $M_2$. The collision-less components
of the galaxies (dark matter and the stars) penetrate through each other freely. Since their mass
far exceeds the gaseous component mass, in the approximate calculation that we perform we may
assume that the masses of the collision-less components of the galaxies are $M_1$ and $M_2$ as
well, and their speeds do not change in the collision. Suppose that the ratios of the galactic ISM
masses to the total masses of the galaxies are $w_1$ and $w_2$, respectively (of course, $w_1\ll
1$, $w_2\ll 1$). Begin from the consideration of a precisely central collision (when the center of
one galaxy passes exactly through the center of another one) for the sake of simplicity. Since the
ISM is not in vacuum state, it undergoes a non-elastic collision, a shock wave occurs between the
gas components. Finally they may merge together (the Bullet cluster gives us an excellent
illustration of how it may happen \cite{bullet2}), and then the final momentum of the cloud is the
sum of the initial momenta of the galactic ISMs. It is convenient to use the ISMs center-of-mass
frame of reference. The initial (and final) galaxy speeds are
\begin{equation}
\label{20b1} v_1=\vcol\frac{w_1 M_1}{w_1 M_1 + w_2 M_2},\qquad v_2=\vcol\frac{w_2 M_2}{w_1 M_1 +
w_2 M_2}
\end{equation}
in this frame, while the resulting gas cloud remains at rest after the collision.

Thus the cloud speeds with respect to the dark matter halos are $v_1$ and $v_2$. Is that enough to
kick the gas out of the the galaxies? It is reasonable to assume for simplicity that $w_1=w_2$.
First we consider the case when the galaxies are approximately equal $M_1=M_2$ and substitute
$\vcol=2000$~{km/s} into (\ref{20b1}). We obtain that $v_1=v_2=1000$~{km/s}, i.e., the gas cloud
after the collision moves with respect to both the galaxies with a speed that is far in excess of
the escape speed even from gigantic galaxies like the Milky Way. Thus the collision can in
principle be strong enough to remove all the gas from both the systems.

Consider the case when $M_1\ll M_2$. Then $v_1\simeq \vcol$, $v_2\simeq \frac{M_1}{M_2} \vcol$. As
we can see, the less massive galaxy still loses all the gas, but the massive component can keep it,
if $v_2$ is less than the escape speed for this galaxy $v_{esc,2}$, i.e., $v_{esc,2}>\vcol
\frac{M_1}{M_2}$. If we accept $v_{esc,2}=400$~{km/s}, $M_2>5 M_1$. So if one of the colliding
galaxies is significantly more massive than the other one, the collision kick itself can be
insufficient to remove gas from the more massive system. However, even in this case an UDG can in
principle be formed as a result of gas evaporation. An upper estimation of the temperature behind
the shock wave is $T_{sw}\sim m_a\vcol^2/2\sim 10^9$~{K}. When the ISMs of the galaxies collide,
the smaller one brings the energy $\sim w_1 M_1 \vcol^2/2$. If we assume that it distributes evenly
over the ISM of the more massive galaxy, the averaged thermal speed of the ISM particles can be
estimated as $v_t \sim \vcol \sqrt{w_1 M_1/w_2 M_2}$. Roughly speaking, gas can evaporate from the
more massive galaxy if $v_t$ exceeds $v_{esc,2}$. Since we assumed that $w_1=w_2$, we finally
obtain the gas evaporation condition
\begin{equation}
\label{20b2}
 v_{esc,2}<v_t\sim \vcol \sqrt{M_1/M_2}.
\end{equation}
As we can see, when the colliding galaxies have significantly different masses, $v_2$ is
proportional to  $M_1/M_2$, while $v_t\propto \sqrt{M_1/M_2}$. Therefore, the gas evaporation is
more effective than the direct kick, if the colliding galaxies differ widely in masses. For the
above-mentioned example of $v_{esc,2}=400$~{km/s}, equation~(\ref{20b2}) gives $M_2>25 M_1$.

Of course, we considered much simplified estimations. First of all, the CM regions may overlap
partially in the collision. A very rough way to consider this situation is to use only the masses
of ISM in the overlapping areas instead of total galactic ISM masses in calculating of $w_1$ and
$w_2$. Then we may substitute these values into (\ref{20b1}). However, this method is quite crude:
apparently, the remainder of the galactic ISM is also perturbed by the collision, even if to a
lesser degree. On the other hand, equation~(\ref{20b2}) for the ISM heating by a smaller colliding
galaxy is too optimistic: the assumption that the kinetic energy of the ISM of the smaller galaxy
distributes over the ISM of the more massive one uniformly is unrealistic. The ISM of the larger
galaxy is the most heated in the collision area, where the gas temperature may approach $T_{sw}$.
However, this very hot substance may be ejected from the galaxy and carry almost all the collision
energy away. It significantly decreases the heating of the rest of the massive galaxy ISM.
Moreover, we ignore the complex structure of ISM: for instance CM consists of individual clouds
separated by WM zones etc.

Complex magnetohydrodynamic simulations is the only way to determine the consequences of a galaxy
collision reliably and quantitatively. However, our simple estimation (\ref{20b1}) is sufficient to
show that the speed the galactic ISM potentially can reach as a result of a galaxy collision
several times exceeds the minimal one necessary to remove the gas away from the system. Therefore,
a total or very significant loss of the gaseous component should occur over a wide range of
collision geometries and galaxy mass ratios, being a frequent consequence of the collisions. The
condition~(\ref{20b2}) of the gas evaporation is even softer, and so the mechanism is even more
effective.

One way or the other, the CM suffers very strong perturbations during the collision, while the gas
removal away from the galaxy is by no means necessary for an UDG formation. Star formation is
possible only in the CM. If the CM is heated and thus transformed into WM or coronal gas, new stars
in the galaxy will not appear anymore until the gas gets cold again. The above-mentioned
alternative mechanisms of the UDG formation (AGN feedback \cite{reines2013}, gas stripping
\cite{yozin2015}, or strong feedback from massive star winds and supernovae \cite{calura2015})
illustrate that an UDG may occur if the star formation in the young galaxy was interrupted by the
CM perturbation. Though the astronomical reason of the perturbation is completely different in our
case, an UDG still may be formed.

To summarize: our estimations show that UDG formation is a quite probable outcome of young galaxy
collisions. An UDG may occur as a result of either a collisional gas removal from the galaxy, or
gas evaporation from the system, or star-forming region heating and destruction. Determination of
exact probability of the UDG formation, as well as relative contributions of collisional removal
and heating effects, require sophisticated and accurate magnetohydrodynamic simulations.

\section{Discussion}
\subsection{The number of UDGs}

Thus, $\sim 300$ central collisions occur in the Coma cluster each $10^9$~years, which corresponds
to $\sim 3000$ collisions in the lifetime of the cluster $\sim 10^{10}$~years. Since two galaxies
are involved into a collision, we conclude that the most of the galaxies composing the Coma cluster
have undergone a central collision. If all of them led to an UDG formation, the UDGs would be
overproduced. It is easy to understand why it does not happen. The majority of the Coma cluster
members are elliptical galaxies. They contain almost no gas, and the star formation there stopped
billions years ago. It is clear that their collisions cannot lead to an UDG formation: enough stars
have already been formed to provide normal surface brightness of these objects. Only the collisions
that occurred during the cluster formation and in $(1- 2)\cdot 10^9$~years after that could lead to
UDG formation. The young galaxies at that epoch contained comparatively few stars and were
extremely rich in gas. Only the dark matter halo and the fairly poor star component remained in the
galaxies after the central collision. In the next $\sim 10^{10}$~years the stellar population got
older and redder, and the galaxies now are red, featureless, and have low surface brightness.
However, the effective radii remain close to that of the galaxies before the collision. Thus
typical UDGs appear. However, even if one of the colliding galaxies contains no gas, we get no UDG:
the gas from another galaxy will not be removed, since it has nothing to collide with.

We may expect $\sim 500 - 600$  central collisions in the Coma cluster in $(1- 2)\cdot 10^9$~years,
and as a result $\sim 1000$ UDGs (among which $\sim 250- 400$ massive) may be formed. This number
is in good agreement with the observational results \cite{koda2015}. In principle, the mechanism
can be even more effective, since we used a rather conservative estimate of $r_1$ and $r_2$.

The mechanism offered in this paper is, in some sense, inevitable. Young galaxies should collide on
the early stages of the Coma cluster's existence, and as our calculation show, even the central
collisions were quite probable, where not only the external hot coronae, but the internal dense gas
regions as well clashed. As a result, the gas necessary for the further star formation was strongly
perturbed and partially removed from the systems. As we can see, theoretically the mechanism is
effective enough to form all the UDGs observed in the Coma cluster. The question of whether the
real UGGs occurred as a result of galaxy collisions, or an another way of formation dominated, is
still an open question.

 The model under consideration predicts that the UDGs space distribution (i.e.,
the number $d$ of the UDGs in a unit volume) roughly follows the density profile $\rho^2(r)$, i.e.,
$n(r)\propto \left((r/r_k)^2+1\right)^{-3}$ for a King model. The corresponding projected density
$n_s$ is $n_s\propto \left((l/r_k)^2+1\right)^{-5/2}$, where $l$ is the projection length of $r$.
The probability $\Psi(\theta)$ to find an UDG in the angular distance interval $[\theta;
\theta+d\theta]$ from the cluster center is
\begin{equation}
\label{20a2}
 \Psi(\theta)\propto \frac{\theta}{\left((\theta/\theta_k)^2+1\right)^{5/2}}
\end{equation}
This equation corresponds the case when the galaxy number density $n(r)$ can be described by the
King model. However, \cite[Fig. 14]{yagi2016} shows that the UDG radial surface number density
tends to track that of bright Coma members, whose surface number density falls as $n_s \sim
\theta^{-2}$ in the outer cluster regions \cite{the1986}, while in the case of the King model it
falls faster $n_s \sim \theta^{-5}$. This discrepancy was not important when we estimated the total
number of collisions in the cluster: the galaxy density density is small at large radius, and the
number of collisions there (proportional to $n^2(r)$) is small.

However, the discrepancy may be important for comparison of the theory predictions with the
observations. In the general case, one should obtain the surface number density $n_s(r)$ from
observations, reconstruct $n(r)$ from these data, calculate $\Psi(\theta)$ and compare it with the
observed surface number density of UDGs. This may allow to recognize the mechanism of UDG
formation, or clarify relative contributions of several possible competitive mechanisms at work. If
observational data follow the distribution of $\Psi(\theta)$ obtained from $n_s(r)$ by the
aforesaid procedure, it would support the central collision mechanism suggested in this paper:
other suppositions about the UDG formation lead to different shapes of $\Psi(\theta)$. It must be
emphasized that a practical implementation of this method is not a simple task. Strictly speaking,
this is the probability of an UDG formation that is proportional to $\rho^2$. Since many cluster
galaxies have radial orbits, the UDG may then move far off the radius where it was formed, smearing
the distribution (\ref{20a2}). Moreover, the history of the cluster formation can be complex. The
problem deserves further investigation.

The perturbations of the globular cluster system around the galaxies by the collision should be
relatively small. This is a natural consequence of the fact that the collision speed far exceeds
the orbital speeds of the globular clusters: the collision time is much shorter than the orbital
ones. As a consequence, the galaxy collision is an adiabatic perturbation with respect to their
globular cluster systems, and the systems are hardly affected by it. Thus, the durability of the
globular cluster systems of the UDGs finds a natural explanation.

Is it possible that, along with the dark matter-dominated objects, the central galaxy collisions
form baryon-dominated galaxies? Unfortunately, it is unlikely. Indeed, in the case of the Bullet
cluster collision one can see a massive gas cloud between the collided clusters, that contains
almost no dark matter. The same cloud (of course, of much lower mass) should occur during the
head-on galaxy collisions. However, the typical collision speed is very large, and the cloud is
heated by the shock wave to the temperature that hardly allows the cloud to be bound by its own
gravity. Therefore, the gas, that is kicked off the galaxies, most likely merely disperses, making
a contribution to the hot halo of the galaxy cluster.

\subsection{The UDG properties}

The suggested model of UDG formation is qualitative. However, we may make several predictions about
UDG properties and compare them with observations. The model under consideration implies that the
UDGs contain little gas and appear from normal galaxies, in which only the very early generations
of stars were formed, and then the star formation was stopped. Indeed, \cite{papastergis2017}
report that, at least, some of UDGs (like R-127-1 or M-161-1) are gas-poor \cite{papastergis2017}.

Recent observations also suggest that the stellar formation in UDGs was interrupted very early.
First, the axial ratio distribution of Coma's UDG stellar components fits prolate rather than
oblate spheroids \cite{burkert2017}. Conceivably young galaxies, at the epoch when most collisions
occur, might likely contain remnant gas reservoirs and be pressure-supported prolate objects.
Second, the radial light distribution of the majority of Coma UDGs is near-exponential,
characteristic of disc galaxies, but not of ellipticals, which follow power laws \cite{yagi2016}.
The exponential component in disc galaxies is formed by the oldest stellar population, the halo
stars. The disc and the bulge are formed later. If gas is removed from the galaxy immediately after
the halo formation, they just do not appear, and we obtain an UDG with the exponential light
distribution. Third, about half of Coma's bright galaxies are radio sources, both thermal and
nonthermal \cite{miller2009}, yet none of its UDGs are radio sources at the detection limit of the
deep VLA survey \cite{struble2017}. In the framework of the model under consideration, this fact is
quite natural. Since the star formation is stopped very early in UDGs, and gas is removed from the
systems, the stellar bulge (or the central dense region in the case of elliptic galaxies) does not
form. As a result, the main engine of active galaxies --- the central supermassive black hole ---
is most likely absent in the systems. Supernovas, young pulsars and other sources of cosmic rays
are few and far between because of extremely low star formation rate. Finally, the poor
interstellar medium cannot hold high energy electrons, even if they are somehow generated. As a
result, the UDGs cannot be radio sources.

However, the above-mentioned prolate shape of UDGs may occur as a result of an another mechanism,
during the binary collision: two initially spherical galaxies should transform into prolate
spheroids with major axes oriented along the collision line as a result of tidal interaction. If
the velocity distribution in the cluster is isotropic, the orientation of major axes of UDGs is
arbitrary. However, as we have already mentioned, the radial motion typically dominates in
clusters, and then the major axes of UDGs should be preferably oriented toward the cluster's
center. \cite{yagi2016} found that this is so indeed for the Coma cluster within $\sim 1$~{Mpc}
from the center. It suggests that the collisions are responsible for the prolate shape of UDGs.

UDGs are found in groups, but are less abundant per unit volume than in clusters
\cite{vanderburg2017}. This observation is also consistent with the collisional mechanism of UDG
formation. Indeed, the velocity dispersion of galaxies are smaller in groups than in clusters, and
so is $\vcol$. Furthermore, the number density $n$ in groups is lower than, at least, in the
central part of clusters. According to equation (\ref{20a1}), both these factors decrease the
number of collisions in a unit volume. Thus, collisions create UDGs in galaxy groups as well, but
less efficiently than in clusters.

Observations \cite{yamanoi2012} indicate that Coma's fainter galaxies are bluer in the cluster's
outskirts and redder in the core region. The mechanism that we suggest is effective only in
clusters or groups: collisions of the field galaxies are relatively rare. On the average, galaxies
in the cluster's center arrive to the cluster and experience the collision leading to an UDG
formation earlier. Therefore, the UDGs in the cluster's center are redder. On the average, the
farther a galaxy from the cluster's center, the later it arrives to the cluster (i.e., to the
region of rich galaxy population). The collision forming a UDG occurs later, and the UDG is on the
average bluer and contains more baryon matter.

Finally, it is quite possible that there are several competitive mechanisms of UDG formation.
Recent observations of isolated UDGs \cite{papastergis2017} show that they can be subdivided into
two groups: gas-rich and gas-poor. The former ones we interpreted by the authors as compatible with
the formation mechanism based on the feedback-driven outflows, but they found puzzling the origin
of the gas-poor UDGs. The mechanism offered in this paper provides a viable explanation for their
formation.

To conclude: of course, the model we use is oversimplified. We apply the model of the Coma cluster
as it is seen now, while the real history of the cluster formation was very complex. The parameters
of the Coma cluster could differ widely from the modern values in $(1 - 2)\cdot 10^9$~years after
initial cluster formation, when the mechanism suggested in this paper was the most effective.
Moreover, galaxies of different masses arrive to the cluster in different ways \cite{balogh2016}.
However, we meaningly set these problems aside. The problem of galaxy cluster formation is very
complex and still far from the complete solution. Any attempt to consider the galaxy collisions in
the framework of a realistic picture of the Coma cluster formation would be cumbersome and
model-dependent. However, the aim of this work is simpler. We just want to show that the galaxy
collisions may lead to the UDG formation and that this process can be quite effective and give an
important contribution to the UDG population of the galaxy clusters. The question of its real
contribution to the recently discovered rich UDG population deserves further consideration.

The work is supported by the Center of Excellence in Astrophysics and Associated Technologies CATA
(PFB06).

\end{document}